\documentclass[final]{arxiv}

\usepackage{graphicx}
\usepackage{natbib}
\begin{document}

\doublespace

\addtolength{\voffset}{0.5in}

\title{\LARGE{Tidal End States of Binary Asteroid Systems}}
\title{\LARGE{with a Nonspherical Component}}
\author{Patrick A. Taylor$^{a,*}$ and Jean-Luc Margot$^{b}$}
\affil{$^{a}$Arecibo Observatory, $^{b}$UCLA}
\email{ptaylor@naic.edu}

\journame{Icarus}
\submitted{22 August 2013}
\revised{22 October 2013}
\accepted{5 November 2013}
\pubonline{14 November 2013}
\pubprint{February 2014 in Volume 229, pp. 418--422}

\pages{13}
\tables{0}
\figures{3}

\quad\newline

\singlespace  

\noindent
NOTICE: this is the author's version of a work that was accepted for publication in Icarus. Changes resulting 
from the publishing process, such as peer review, editing, corrections, structural formatting, and other 
quality control mechanisms may not be reflected in this document. Changes may have been made to this work 
since it was submitted for publication.  A definitive version was subsequently published in Icarus 229, 418--422, 
February 2014, DOI: 10.1016/j.icarus.2013.11.008.\\
\\
Publisher's copy:  http://dx.doi.org/10.1016/j.icarus.2013.11.008\\
\\
Free access to publisher's copy (until 26 March 2014):  http://elsarticle.com/1ejGrKG\\ 

\quad\clearpage

\addtolength{\voffset}{-0.5in}


\noindent ABSTRACT:  
We derive the locations of the fully synchronous end states of tidal evolution for binary asteroid systems having 
one spherical component and one oblate- or prolate-spheroid component. Departures from a spherical shape, 
at levels observed among binary asteroids, can result in the lack of a stable tidal end state for particular combinations
of the system mass fraction and angular momentum, in which case the binary must collapse to contact.  We 
illustrate our analytical results with near-Earth asteroids (8567) 1996 HW$_{1}$, (66391) 1999 KW$_{4}$, and 
69230 Hermes.


\section{Introduction}
\label{sec:intro}
Recent studies have examined energy, stability, and orbital relative equilibria in the planar two-body problem for a 
non-rotating sphere and an arbitrary, rotating ellipsoid~\citep{sche07,bell08} and approximately for two arbitrary, 
rotating ellipsoids~\citep{sche09}.  Here, we examine the special case of a rotating sphere interacting with a rotating 
oblate or prolate spheroid and provide exact, tractable analytical results for the locations of the fully synchronous 
end states of tidal evolution.  The terms fully synchronous tidal end state and orbital relative equilibrium can be used 
interchangeably to describe a zero-eccentricity binary system that has ceased tidally evolving because the spin 
rates of both components have synchronized to the mean motion of the components about the center of mass of the 
system.  

This note is organized as follows.  In Section 2, we review fully synchronous tidal end states of a binary system 
consisting of two spheres.  Section 3 extends the discussion to a sphere interacting with an ellipsoid and explores 
the specific cases of oblate and prolate spheroids with applications to real asteroid systems.  Comparisons to previous 
work in Sections 3 and 4 place this work in context and possible avenues for contact-binary formation are suggested.


\section{Fully synchronous orbits with spherical components}
\label{sec:spherical}

The locations of the fully synchronous end states of tidal evolution for binary asteroids with spherical components were 
discussed by~\citet{tayl11material} and are summarized here.  For components of equal, uniform density $\rho$ with 
radii $R_{1}$ and $R_{2}$ and mass ratio $q=M_{2}/M_{1}=\left(R_{2}/R_{1}\right)^3$ separated by a distance $a$ in 
their circular mutual orbit, the sum of the orbital and spin angular momentum $J$ upon full synchronization, scaled by 
$J^{\prime}=\sqrt{G\left(M_{1}+M_{2}\right)^{3}R_{\rm eff}}$, where $R_{\rm eff}$ is the effective radius of a sphere 
with the same volume as both components combined, is:
\begin{equation}
\frac{J}{J^{\prime}}~=~\frac{q}{\left(1+q\right)^{13/6}}\,\left(\frac{a}{R_{1}}\right)^{1/2}+\frac{2}{5}\,\frac{1+q^{5/3}}{\left(1+q\right)^{7/6}}\,\left(\frac{a}{R_{1}}\right)^{-3/2}
\label{eq:jcontour}
\end{equation}
\noindent
[cf.~\citet{tayl11material}, Eq. (8)].  The term on the left, proportional to $a^{1/2}$, is the orbital angular momentum of the 
system revolving with mean motion $n$, given by Kepler's Third Law, scaled by $J^{\prime}$.  The term on the right, 
proportional to $a^{-3/2}$, is the spin angular momentum of the two components, both rotating with spin rate $n$, scaled 
by $J^{\prime}$.  The $1+q^{5/3}$ term is proportional to the sum of the moments of inertia of the two bodies; removing the 
$q^{5/3}$ term amounts to ignoring the spin angular momentum of component 2.  Depending on the mass ratio and the 
total angular momentum of the system, Eq.~(\ref{eq:jcontour}) may have zero, one (degenerate), or two solutions (one 
unstable and one stable), corresponding to the number of fully synchronous orbits supported by the system.  The total 
energy when the system has fully synchronized may be positive or negative depending on the parameters of the system.  
One can show that the zero-energy limit always falls within the stability limit that splits the unstable and stable solutions 
such that all stable, fully synchronous orbits have negative energy, \textit{i.e.}, they are gravitationally bound.

\begin{figure}[!p]
\begin{center}
\includegraphics[angle=180., scale=0.5]{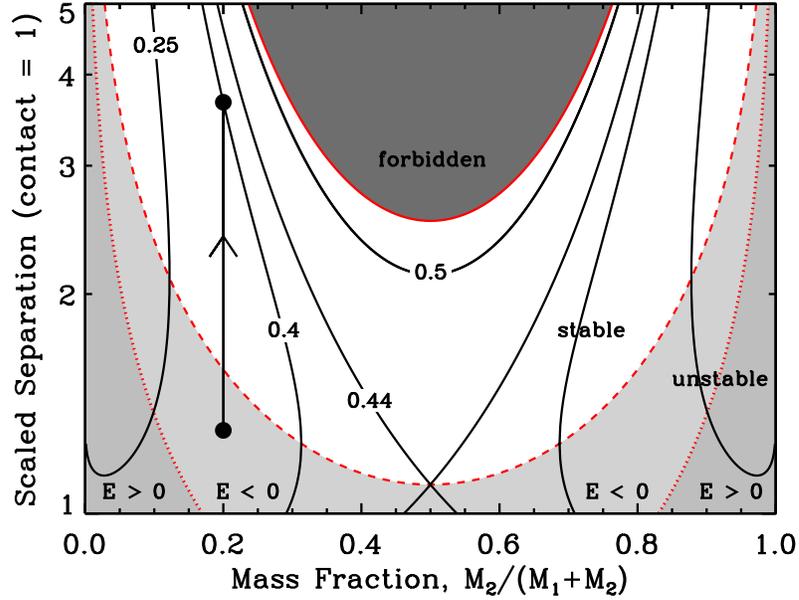}
\caption[Synchronous orbits for the two-sphere case]{
Component separation $a$, scaled to the contact limit, for the fully synchronous orbits of a two-sphere binary system with 
mass fraction $v$ and angular momentum $J/J^{\prime}$.  The black curves indicate the inner (when not within contact limit) 
and outer synchronous orbits for $J/J^{\prime}=0.25, 0.4, 0.44$ and $0.5$.  The red dotted curve is the zero-energy limit; tidal 
end states above this limit have negative energy ($E < 0$) and must remain bound.  The red dashed curve is the stability limit 
that splits the unstable inner orbits in the gray regions from the stable outer orbits in the white region.  The darkest region 
above the solid red line represents the angular-momentum limit for $J/J^{\prime}=0.5$ and is inaccessible to systems with 
$J/J^{\prime} \le 0.5$.  See Section~\ref{sec:limits} and~\citet{tayl11material} for details on these limits.  A binary system tidally 
evolves upward along a vertical line at mass fraction $v$, away from the gray regions and into the white region, where it reaches 
the stable outer synchronous orbit at the intersection with its corresponding $J/J^{\prime}$ contour.  A $v=0.2$ $(q=0.25)$ 
binary system with $J/J^{\prime}=0.4$ is shown evolving from a state initially near contact.}
\label{fig:spheres}
\end{center}
\end{figure}

For plotting purposes, we transform from mass ratio $q$ to mass fraction $v=M_{2}/\left(M_{1}+M_{2}\right)$ and scale 
the separation $a$ by $R_{1}+R_{2}$, the contact limit.  Figure~\ref{fig:spheres} shows, for a two-sphere binary system, 
the locations of the fully synchronous orbits using contours of angular momentum $J/J^{\prime}$.  Because the 
components are similar in shape, the diagram is mirror symmetric about $v=0.5$; this will not be the case when one 
component is nonspherical.  Unstable inner synchronous orbits, the solutions below the stability limit in Fig.~\ref{fig:spheres}, 
almost always fall within the contact limit, with the exception of the $J/J^{\prime} = 0.25$ curve, similar to the angular 
momentum found in most large main-belt binary systems likely formed by collisions.  In systems with $J/J^{\prime} \sim 0.4$, 
similar to near-Earth binaries and small main-belt binaries likely formed via spin-up processes, the secondary is formed 
beyond the inner synchronous orbit and will naturally tidally evolve outward, vertically through the diagram, until reaching 
the outer synchronous orbit at the intersection with its corresponding angular-momentum contour.  Of course, this is a 
simplistic view because the post-fission dynamical environment of a newly formed binary asteroid is chaotic~\citep{jaco11}, 
carrying the risk of ejection or re-impact of the secondary or the secondary itself undergoing fission.  Once the system has
settled, the steady, comparatively quiescent, tidal evolution to the outer synchronous orbit continues as in Fig.~\ref{fig:spheres}.  
An equal-mass binary with $v=0.5$ must have $J/J^{\prime} > 0.44$ (more exactly, 0.43956) to have a stable tidal end state.

\section{Fully synchronous orbits with a nonspherical component}
\label{sec:nonspherical}

Let component 1 of the binary system be a uniform-density ellipsoid with principal semi-axes $a_{0} \ge a_{1} \ge a_{2}$ 
such that the equivalent radius of the ellipsoid is $R_{1}=(a_{0}a_{1}a_{2})^{1/3}$.  For rotation about 
the shortest principal axis, the ratio of the moment of inertia of the ellipsoid to that of its equivalent-volume sphere with 
radius $R_{1}$ is the nonsphericity parameter~\citep{desc08}: 
\begin{equation}
\lambda~=~\frac{1+\beta^{2}}{2\left(\alpha\beta\right)^{2/3}},
\end{equation}
\noindent
where $\alpha=a_{2}/a_{0}$, $\beta=a_{1}/a_{0}$, and $\alpha\le\beta\le1$.  The nonsphericity parameter 
is always larger than unity because any departure from a spherical shape requires displacing mass farther 
from the spin axis and increases the moment of inertia of the body.  Component 2 is assumed to remain 
spherical.
To retain orbital relative equilibrium, the sphere must orbit above one of the principal axes of the ellipsoid and 
the system must rotate about another principal axis of the ellipsoid at a specific rate~\citep{sche06} given by: 
\begin{equation}
n^{2}~=~\frac{3}{2}\,G\left(M_{\rm 1}+M_{\rm 2}\right)\int_{r^{2}-a_{i}^{2}}^{\infty}\frac{{\rm d}u}{\left(a_{i}^{2}+u\right)\Delta(u)}
\label{eq:n2full}
\end{equation}
\noindent
[cf.~\citet{sche07}, Eq. (18)], where $\displaystyle{\Delta(u)=\sqrt{\left(a_{0}^{2}+u\right)\left(a_{1}^{2}+u\right)\left(a_{2}^{2}+u\right)}}$, $a_{i}$ 
is the principal semi-axis that the sphere orbits above, and $r$ is the orbital separation of the bodies ($r$ is the 
semimajor axis $a$ for the circular orbits considered here).  Defining $\bar{a}=a/a_{0}$ and $u^{\prime}=u/a_{0}^{2}$, 
the mean motion becomes:
\begin{equation}
n^{2}~=~\frac{G\left(M_{1}+M_{2}\right)}{a^{3}}\,f\left(\alpha,\beta,\bar{a},a_{i}\right),
\label{eq:n2f}
\end{equation}
\noindent
introducing $f$ as the dimensionless integral:
\begin{equation}
f\left(\alpha,\beta,\bar{a},a_{i}\right)~=~\frac{3}{2}\,\bar{a}^{3}\,\int_{\bar{a}^{2}-\left(\frac{a_{i}}{a_{0}}\right)^{2}}^{\infty}\frac{{\rm d}u^{\prime}}{\left(\left(\frac{a_{i}}{a_{0}}\right)^{2}+u^{\prime}\right)\sqrt{\left(1+u^{\prime}\right)\left(\alpha^{2}+u^{\prime}\right)\left(\beta^{2}+u^{\prime}\right)}}.
\label{eq:f}
\end{equation}
\noindent
For two spheres, Eq.~(\ref{eq:n2f}) simplifies to Kepler's Third Law as the integral $f$ goes to unity.  To apply this 
condition to an ensemble of systems while accounting for the spins of both components and the orbital mean motion, 
we use a dimensionless form of the angular momentum that is applicable to binary systems with any absolute size, 
mass, and separation.  Starting from Eq.~(\ref{eq:jcontour}), when component 1 is nonspherical, the effective radius 
$R_{1}$ is by definition $\left(\alpha\beta\right)^{1/3}a_{0}$, the contribution of the (scaled) moments of inertia of the 
two components to the spin angular momentum increases from $1+q^{5/3}$ to $\lambda+q^{5/3}$, and the mean motion 
$n$ includes the additional factor of $f\left(\alpha, \beta, \bar{a}, a_{i}\right)^{1/2}$ compared to the two-sphere case.  
Upon simplification, the total angular momentum $J/J^{\prime}$ of a sphere and ellipsoid in a fully synchronous orbit 
satisfies:
\begin{equation}
\frac{J}{J^{\prime}}~=~\left(\alpha\beta\right)^{-1/6}\,\left[\frac{q}{\left(1+q\right)^{13/6}}\,\bar{a}^{1/2}+\frac{2}{5}\,\frac{1}{\left(1+q\right)^{7/6}}\left(\frac{1+\beta^{2}}{2}+\left(\alpha\beta\right)^{2/3}q^{5/3}\right)\bar{a}^{-3/2}\right]\left[f\left(\alpha,\beta,\bar{a},a_{i}\right)\right]^{1/2},
\label{eq:jcontournon}
\end{equation}
\noindent
recalling that $\bar{a}=a/a_{0}$.  In the limit that the nonspherical component approaches a sphere, $\alpha$, $\beta$, 
and the integral $f$ go to unity and $\bar{a}$ is equivalent to $a/R_{1}$, which recovers the two-sphere case of 
Eq.~(\ref{eq:jcontour}) explored by~\citet{tayl11material} and shown in Fig.~\ref{fig:spheres}.  


\subsection{Angular-momentum, stability, and zero-energy limits}
\label{sec:limits}

Three dynamical limits:  the angular-momentum limit, the stability limit, and the zero-energy limit, break up the 
parameter space of mass fraction and separation, and all three depend on the shape of the nonspherical 
component of the binary system.  The angular-momentum limit follows from Eq.~(\ref{eq:jcontournon}) by setting 
the spin angular momentum (the term proportional to $\bar{a}^{-3/2}$) to zero and rearranging such that the 
maximum separation of the components $\bar{a}_{\rm max}=a_{\rm max}/a_{0}$ for a given angular momentum 
$J/J^{\prime}$ is the numerical solution to:
\begin{equation}
\bar{a}_{\rm max}\,f\left(\alpha,\beta,\bar{a}_{\rm max},a_{i}\right)~=~\left(\alpha\beta\right)^{1/3}\,\frac{\left(1+q\right)^{13/3}}{q^{2}}\,\left(J/J^{\prime}\right)^{2}. 
\label{eq:angmomlimitnon}
\end{equation}
\noindent
In the limit that the nonspherical shape approaches a sphere, $\alpha$, $\beta$, and $f$ go to unity, reproducing 
the two-sphere result [cf.~\citet{tayl11material}, Eq. (5)].  The stability limit $\bar{a}_{\rm stab}$, which splits the 
solutions to Eq.~(\ref{eq:jcontournon}) into unstable inner and stable outer orbits, is given by the root of:
\begin{equation}
\frac{{\rm d}}{{\rm d}\bar{a}}\left[\frac{J}{J^{\prime}}\left(\alpha,\beta,\bar{a},a_{i}\right)\right]~=~0.
\end{equation}
\noindent
Due to the complicated dependence of the integral in Eq.~(\ref{eq:jcontournon}) on separation, the stability limit is 
not algebraic as in the two-sphere case [cf.~\citet{tayl11material}, Eq. (11)] and is left to a numerical solution.  

The total energy of a binary system is given by the sum of the rotational, orbital, and gravitational potential energies.  
When the components are fully synchronized to the orbital mean motion $n$:
\begin{equation}
E~=~\frac{1}{2}\frac{M_{1}M_{2}}{M_{1}+M_{2}}\,a^{2}n^{2}+\frac{1}{2}I_{1}n^{2}+\frac{1}{2}I_{2}n^{2}+V,
\end{equation}
\noindent
where the rotational energy is in terms of the polar moments of inertia $I$ of the components, $\left(2/5\right)MR^{2}$ 
for a sphere and a factor of $\lambda$ greater for an ellipsoid.  The gravitational potential energy 
$V$~\citep[\textit{e.g.},][and references therein]{sche94} can be written as:
\begin{equation}
V~=~-\frac{GM_{1}M_{2}}{a}\left[\frac{3}{4}\,\bar{a}\int_{\bar{a}^{2}-\left(\frac{a_{i}}{a_{0}}\right)^{2}}^{\infty}\left(1-\frac{\bar{a}^{2}}{\left(\frac{a_{i}}{a_{0}}\right)^{2}+u^{\prime}}\right)\frac{{\rm d}u^{\prime}}{\sqrt{\left(1+u^{\prime}\right)\left(\alpha^2+u^{\prime}\right)\left(\beta^2+u^{\prime}\right)}}\right]
\label{eq:V}
\end{equation}
\noindent
when the sphere orbits above semi-axis $a_{i}$ of the ellipsoid.  Note the familiar result for the gravitational potential 
energy between two point masses, which is reproduced when the term in brackets goes to unity for two spheres.  
One can split the integral for $V$ into two terms and define another dimensionless integral:
\begin{equation}
g\left(\alpha,\beta,\bar{a},a_{i}\right)~=~\frac{3}{2}\,\bar{a}\,\int_{\bar{a}^{2}-\left(\frac{a_{i}}{a_{0}}\right)^{2}}^{\infty}\frac{{\rm d}u^{\prime}}{\sqrt{\left(1+u^{\prime}\right)\left(\alpha^{2}+u^{\prime}\right)\left(\beta^{2}+u^{\prime}\right)}}
\end{equation}
\noindent
similar to $f$.  Then, for the instance when the total energy $E$ is zero, after applying Eq.~(\ref{eq:n2f}), the 
critical separation $\bar{a}_{\rm E}$ satisfies:
\begin{equation}
\bar{a}_{\rm E}\left[\frac{g\left(\alpha,\beta,\bar{a}_{\rm E},a_{i}\right)}{f\left(\alpha,\beta,\bar{a}_{\rm E},a_{i}\right)}-2\right]^{1/2}~=~\left[\frac{2}{5}\,\frac{1+q}{q}\left(\frac{1+\beta^{2}}{2}+\left(\alpha\beta\right)^{2/3}q^{5/3}\right)\right]^{1/2},
\label{eq:energy}
\end{equation}
\noindent
which recovers the two-sphere, zero-energy limit [cf.~\citet{tayl11material}, Eq. (15)] as $\alpha$ and $\beta$ 
go to unity and $g/f\rightarrow 3$.


\subsection{Comparison to previous work}
\label{sec:bellerose}

\citet{sche07} and~\citet{bell08} analyze a planar two-body problem consisting of a triaxial ellipsoid and a sphere 
in mutual orbit.  Concentrating on~\citet{bell08}, our analysis follows similarly, but with two key differences.  First, we 
normalize the total angular momentum as $J/J^{\prime}$, while they normalize their angular momentum $K$ 
[cf.~\citet{bell08}, Eq. (18)] such that:
\begin{equation}
J/J^{\prime}~=~\left(\alpha\beta\right)^{-1/6}\,v\left(1-v\right)^{7/6}\,K.
\label{eq:ktoj}
\end{equation}
\noindent
The second important difference is that we account for the spin angular momentum of the sphere, the $q^{5/3}$ term 
in Eqs.~(1) and (6), which makes a non-negligible contribution in equal-mass binaries and when the sphere is the 
primary component.  As a result, Eq.~(\ref{eq:ktoj}) must be supplemented by the spin angular momentum of the 
sphere scaled by $J^{\prime}$.  \citet{sche07} also ignores the spin angular momentum of the sphere in orbit because, 
under ideal conditions, a perfect sphere in orbit cannot be tidally torqued or transfer angular momentum, but we consider 
the sphere to have a slight tidal or permanent deformation to allow for spin-orbit coupling even though it is treated 
mathematically as a sphere.  Combining these two points prevents a one-to-one mapping between the results of this 
work for a value of $J/J^{\prime}$ and those of~\citet{bell08} for a value of $K$, though the curves that trace out our 
fully synchronous orbits and their orbital relative equilibria, as well as the stability and zero-energy limits, have the same 
inherent meanings.  The difference between these approaches is evident when comparing the shapes of the curves 
in our Fig.~\ref{fig:spheres} to those in Figs.~7 and 8 of~\citet{sche07} and Fig.~6 of~\citet{bell08}.

\citet{sche09} considers binary systems of two triaxial ellipsoids and expands the gravitational potential between 
the components to second order in the moments of inertia.  In this work, we use the exact form of the gravitational 
potential in Eq.~(\ref{eq:V}), which leads to the equilibrium condition in Eq.~(\ref{eq:n2full}).  In the limiting cases 
of spheres interacting with oblate and prolate spheroids, we can test the accuracy of the~\citet{sche09} method.  
The key difference between our end states and the equilibria of~\citet{sche09} is that we use a single value of 
angular momentum for all mass fractions, while, for each mass fraction,~\citet{sche09} uses the specific value 
of angular momentum that allows the components to fission from contact.  In other words,~\citet{sche09} uses the 
angular-momentum value that has an unstable inner synchronous orbit at the contact limit for that specific mass 
fraction, which prevents a direct one-to-one mapping of our results to theirs.


\subsection{Oblate component, $a_{2}<a_{1}=a_{0}$}
\label{sec:oblate}

\begin{figure}[!p]
\begin{center}
\includegraphics[angle=180., scale=0.5]{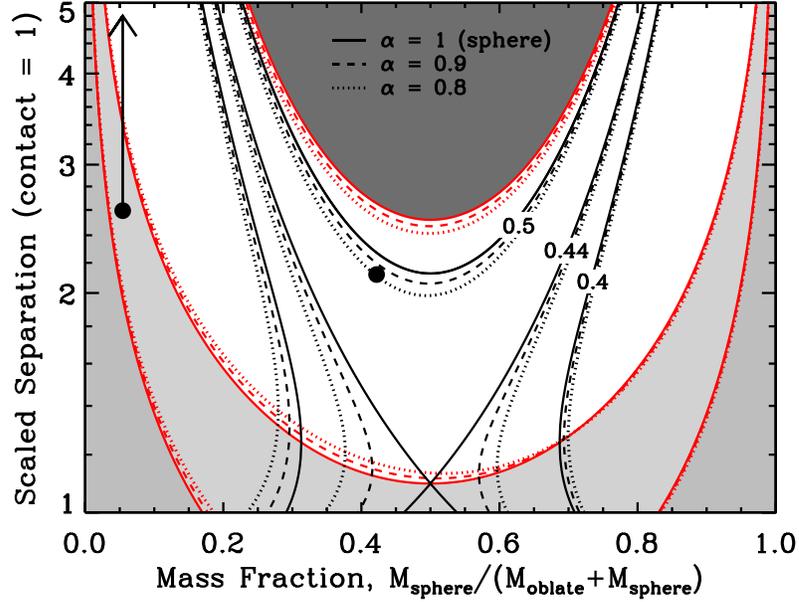}
\caption[Synchronous orbits for an oblate shapes with $\alpha=0.9$ and $\alpha=0.8$]{
Component separation $a$, scaled to the contact limit, for the fully synchronous orbits of a sphere/oblate-spheroid 
binary system with oblateness $\alpha=$~1, 0.9, and 0.8, mass fraction $v$, and angular momentum 
$J/J^{\prime} = $~0.4, 0.44, and 0.5.  Solid lines correspond to the $\alpha=1$ (two-sphere) case, dashed lines 
correspond to $\alpha=0.9$, and dotted lines correspond to $\alpha=0.8$.  Red curves represent the angular-momentum, 
stability, and zero-energy limits as in Fig.~\ref{fig:spheres}.  Further annotations are suppressed for clarity.  The gray 
regions are shaded with respect to the $\alpha=1$ case, but could be shaded for any value of $\alpha$.  The darkest 
region is inaccessible to systems with $J/J^{\prime} \le 0.5$.  The current state of binary near-Earth asteroid (66391) 
1999 KW$_{4}$, with oblateness $\alpha=0.85$, is the circle at left.  It will continue to tidally evolve vertically through 
the diagram until intersecting with the $J/J^{\prime}=0.4$ curve.  At center, nearly equal-mass binary 69230 Hermes, 
with oblateness $\alpha=0.8$, has reached a stable, fully synchronous state with $J/J^{\prime}\sim0.5$.}
\label{fig:kw4}
\end{center}
\end{figure}

In near-Earth binary asteroid systems, the rapid rotation of the primary component tends to produce an oblate 
shape with loose regolith built up in a circular equatorial belt, \textit{e.g.}, the primary component of (66391) 
1999 KW$_{4}$~\citep{ostr06}.  For an oblate spheroid with identical equatorial principal axes rotating 
about the shortest principal axis, the semi-axes $a_{2}<a_{1}=a_{0}$ ($\alpha<\beta=1$) 
and $\lambda=\alpha^{-2/3}$.  For the sphere along $a_{0}$ or equivalently anywhere in the equator 
plane of the oblate component, the dimensionless integral $f$ from Eq.~(\ref{eq:f}) simplifies to:
\begin{equation}
f\left(\alpha,1,\bar{a},a_{0}\right)~=~\frac{3}{2}\,\bar{a}^{3}\int_{\bar{a}^2-1}^{\infty}\frac{{\rm d}u^{\prime}}{\left(1+u^{\prime}\right)^{2}\left(\alpha^{2}+u^{\prime}\right)^{1/2}}.
\label{eq:foblate}
\end{equation}
\noindent
Evaluating Eq.~(\ref{eq:foblate}) and substituting into Eq.~(\ref{eq:n2f}), the necessary spin rate for orbital 
relative equilibrium, in terms of the uniform density $\rho$ of the components, is:
\begin{equation}
n^{2}~=~2\pi\,G\rho\,\left(1+q\right)\,\frac{\alpha}{1-\alpha^{2}}\left[\frac{1}{\sqrt{1-\alpha^{2}}}\,\tan^{-1}\left(\sqrt{\frac{1-\alpha^{2}}{\bar{a}^{2}+\alpha^{2}-1}}\right)-\frac{1}{\bar{a}^{2}}\sqrt{\bar{a}^{2}+\alpha^{2}-1}\right],
\end{equation}
\noindent
recalling that $\bar{a}=a/a_{0}$ and $q=M_{\rm sphere}/M_{\rm oblate}$.  The fully synchronous orbits 
then satisfy the angular-momentum equation from Eq.~(\ref{eq:jcontournon}):
\begin{eqnarray}
J/J^{\prime} & = & \left[\frac{q}{\left(1+q\right)^{13/6}}\,\bar{a}^{2}+\frac{2}{5}\frac{1}{\left(1+q\right)^{7/6}}\left(1+\alpha^{2/3}q^{5/3}\right)\right]\\
 & & \quad \times \left[\frac{3/2}{\alpha^{1/3}\left(1-\alpha^{2}\right)}\right]^{1/2}\left[\frac{1}{\sqrt{1-\alpha^{2}}}\,\tan^{-1}\left(\sqrt{\frac{1-\alpha^{2}}{\bar{a}^{2}+\alpha^{2}-1}}\right)-\frac{1}{\bar{a}^{2}}\sqrt{\bar{a}^{2}+\alpha^{2}-1}\right]^{1/2}. \nonumber
\label{eq:jcontouroblate}
\end{eqnarray}  
\noindent
The above expression describes the synchronous orbits as contours of constant $J/J^{\prime}$ for a 
sphere/oblate-spheroid system in the same way as Eq.~(\ref{eq:jcontour}) does for a two-sphere system.  Applying 
the transformations $q=v/\left(1-v\right)$ and $\bar{a}=\left(1+\alpha^{1/3}q^{1/3}\right)\,\left[a/(a_{0}+R_{2})\right]$ 
to arrive at mass fraction and separation scaled to the contact limit, the angular-momentum contours are shown in 
Fig.~\ref{fig:kw4}.

The curves in Fig.~\ref{fig:kw4} are nearly symmetric about $v=0.5$ (equal masses) with the effect of $\alpha$ 
subdued for large mass fractions where the smaller secondary is oblate rather than the larger primary component.  
At small mass fractions, such as that of 1999 KW$_{4}$, a typical near-Earth binary with $\alpha=0.85$, the effect 
of primary oblateness on the curves in Fig.~\ref{fig:kw4} is barely perceptible.  At nearly equal masses, it is clearer 
that oblateness, for a given mass fraction and angular momentum, causes the inner synchronous orbit to push 
outward and the outer synchronous orbit to push inward by several percent, while the size of the largest supportable 
companion decreases (the lack of solutions near $v=0.5$).  For instance, while an angular-momentum budget of 
$J/J^{\prime}=0.44$ can support any two-sphere binary system, having a component with oblateness of $\alpha=0.8$ 
would result in collapse to a contact binary for $v=0.4-0.6$ as a stable, fully synchronous orbit no longer exists to 
support such a system.  A system having an oblate component with $\alpha=0.8$, such as 69230 Hermes~\citep{marg06iau}, 
requires an increase in total angular momentum of at least 3.2\% to account for the larger moment of inertia of the 
nonspherical component and support a spherical companion of any mass fraction.  Hermes though, with an adequate 
angular-momentum budget of $J/J^{\prime}\sim0.5$, is not in danger of collapsing to a contact binary and has 
reached a stable tidal end state with a separation of roughly twice the contact limit.

\subsection{Prolate component, $a_{2}=a_{1}<a_{0}$}
\label{sec:prolate}

\begin{figure}[!p]
\begin{center}
\includegraphics[angle=180., scale=0.5]{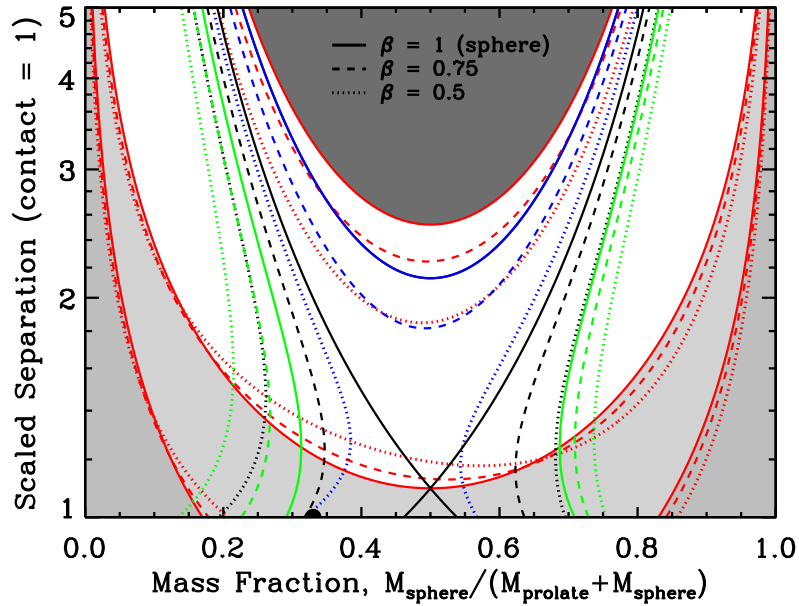}
\caption[Synchronous orbits for a prolate shape with $\beta=0.75$ and $\beta=0.5$]{
Component separation $a$, scaled to the contact limit, for the fully synchronous orbits of a sphere/prolate-spheroid 
binary system with prolateness $\beta=$~1, 0.75, and 0.5, mass fraction $v$, and angular momentum 
$J/J^{\prime} = $~0.4, 0.44, and 0.5, colored green, black, and blue, respectively.  The sphere orbits above the long 
axis of the prolate spheroid.  Solid lines correspond to the $\beta=1$ (two-sphere) case, dashed lines correspond to 
$\beta=0.75$, and dotted lines correspond to $\beta=0.5$.  Red curves represent the angular-momentum, stability, 
and zero-energy limits as in Fig.~\ref{fig:spheres}.  Further annotations are suppressed for clarity.  The gray regions 
are shaded with respect to the $\beta=1$ case, but could be shaded for any value of $\beta$.  The darkest region is 
inaccessible to systems with $J/J^{\prime} \le 0.5$.  The circle on the horizontal axis at $v=0.33$ is contact binary 
(8567) 1996 HW$_{1}$ with $\beta=0.6$.}
\label{fig:prolong}
\end{center}
\end{figure}

For a prolate spheroid with two equivalent shorter principal semi-axes, one of which is aligned with the 
spin axis, $a_{2}=a_{1}<a_{0}$ ($\alpha=\beta<1$) and $\lambda=\left(1+\beta^{2}\right)/\left(2\beta^{4/3}\right)$.  
Here, we only consider the sphere orbiting above the long axis.  Although the intermediate-axis case follows 
similarly mathematically, \citet{sche06} has shown the intermediate-axis case is never energetically stable.  
When the spherical component orbits above the longest principal semi-axis $a_{0}$, the dimensionless integral 
$f$ from Eq.~(\ref{eq:f}) for a sphere/prolate-spheroid binary simplifies to:
\begin{equation}
f\left(\beta,\beta,\bar{a},a_{0}\right)~=~\frac{3}{2}\,\bar{a}^{3}\int_{\bar{a}^{2}-1}^{\infty}\frac{{\rm d}u^{\prime}}{\left(1+u^{\prime}\right)^{3/2}\left(\beta^{2}+u^{\prime}\right)}.
\label{eq:fprolate}
\end{equation}
\noindent
Evaluating Eq.~(\ref{eq:fprolate}) and substituting into Eq.~(\ref{eq:n2f}), the necessary spin rate for orbital 
relative equilibrium, in terms of the uniform density $\rho$ of the components, is:
\begin{equation}
n^{2}~=~4\pi\,G\rho\,\left(1+q\right)\,\frac{\beta^{2}}{1-\beta^{2}}\left[\frac{1}{\sqrt{1-\beta^{2}}}\,\tanh^{-1}\left(\frac{\sqrt{1-\beta^{2}}}{\bar{a}}\right)-\frac{1}{\bar{a}}\right],
\label{eq:n2prolate}
\end{equation}
\noindent
recalling that $\bar{a}=a/a_{0}$ and $q=M_{\rm sphere}/M_{\rm prolate}$.  The fully synchronous orbits 
then satisfy the angular-momentum equation from Eq.~(\ref{eq:jcontournon}):
\begin{eqnarray}
\label{eq:jcontourprolate1}
J/J^{\prime} & = & \left[\frac{q}{\left(1+q\right)^{13/6}}\,\bar{a}^{2}+\frac{2}{5}\,\frac{1}{\left(1+q\right)^{7/6}}\left(\frac{1+\beta^{2}}{2}+\beta^{4/3}q^{5/3}\right)\right]\\
 & & \quad \times \left[\frac{3}{\beta^{2/3}\left(1-\beta^{2}\right)}\right]^{1/2}\left[\frac{1}{\sqrt{1-\beta^{2}}}\,\tanh^{-1}\left(\frac{\sqrt{1-\beta^{2}}}{\bar{a}}\right)-\frac{1}{\bar{a}}\right]^{1/2}. \nonumber
\end{eqnarray}
\noindent
The above expression describes the synchronous orbits as contours of constant $J/J^{\prime}$ for a 
sphere/prolate-spheroid system in the same way as Eq.~(\ref{eq:jcontour}) does for a two-sphere system.  Applying 
the transformations $q=v/\left(1-v\right)$ and $\bar{a}=\left(1+\beta^{2/3}q^{1/3}\right)\,\left[a/(a_{0}+R_{2})\right]$ 
to arrive at mass fraction and separation scaled to the contact limit, the angular-momentum contours are shown in 
Fig.~\ref{fig:prolong}.

The asymmetry across $v=0.5$ (equal masses) is clearer in the prolate case primarily due to the wider range of $\beta$ 
values used based on observed nonspherical asteroid shapes.  Similar to the oblate case, the effect of $\beta$ is more 
subdued for large mass fractions, where the secondary is prolate rather than the larger primary component, and most 
important for nearly equal-mass components.  Departure from a spherical shape causes the inner synchronous orbit to 
push outward and the outer synchronous orbit to push inward by as much as tens of percent depending on the mass 
fraction and degree of prolateness of the nonspherical component, while the size of the largest supportable companion 
decreases (the lack of solutions near $v=0.5$).  An angular-momentum budget of $J/J^{\prime}=0.44$ can support any 
two-sphere binary system, but a component with prolateness $\beta=0.75$ would result in collapse to a contact binary 
for $v=0.35-0.6$ and roughly $v=0.25-0.7$ for $\beta=0.5$.  To support a spherical companion of any mass fraction, 
systems with $\beta=0.75$ and 0.5 require substantial increases in total angular momentum of at least 5.6\% and 16\%, 
respectively.  As an example, we approximate contact binary (8567) 1996 HW$_{1}$ as a $\beta=0.6$ prolate spheroid 
with a $v=0.33$ sphere at the end of its long axis~\citep{magr11}.  Using Eq.~(\ref{eq:jcontourprolate1}), to remain in 
contact, $J/J^{\prime}$ must be less than 0.475, which by Eq.~(\ref{eq:n2prolate}) corresponds to a density $\rho$ 
greater than $\sim$0.85 g cm$^{-3}$ given its present rotation rate.  The density constraint for $J/J^{\prime}=0.475$ 
is 15\% less when using the true shape of 1996 HW$_{1}$ and would be more stringent for a faster rotation period.

\section{Discussion}
\label{sec:disc}

We have presented analytical formulae for contours of constant angular momentum for ensembles of binary systems 
consisting of an oblate/prolate component and a spherical component that, in turn, determine the fully synchronous 
tidal end states for specific binary systems according to their mass fractions.  In retaining the spin angular momentum 
of the sphere, we extend the results of~\citet{bell08}.  In general, the presence of a nonspherical component breaks the 
symmetry about $v=0.5$ of a two-sphere system and, for comparable angular momenta, shifts inner synchronous orbits 
outward and outer synchronous orbits inward and reduces the size of the largest supportable companion.  At disparate 
mass ratios though, like those of typical near-Earth binaries, the effect of a nonspherical component on the locations of 
the fully synchronous tidal end states is minimal, especially given the extreme timescales required to reach them~\citep{tayl11material}.  
The strongest effect is in the nearly equal-mass regime ($v\sim0.5$), where tidal timescales are much less than the dynamical 
lifetimes of asteroids.  To support the same mass fraction as a two-sphere binary, the system requires a larger injection 
of angular momentum during binary formation to overcome the increased moment of inertia of the nonspherical primary.  
Once a nearly equal-mass binary is formed, and survives against re-impact while tidally evolving through the dynamically 
unstable region within the stability limit~\citep{sche09,jaco11}, two avenues allow collapse to a contact binary:  loss of 
enough angular momentum, \textit{e.g.}, through YORP and/or BYORP thermal torques, that a stable tidal end state no 
longer exists for the system, or one unrelaxed component deforms to a more oblate or prolate shape that cannot support 
the binary with the existing angular-momentum budget.  These avenues are in addition to other proposed contact-binary 
formation mechanisms such as a secondary-fission event resulting in the gentle impact of a smaller component onto the 
primary component~\citep{jaco11} or the gravitational reaccumulation of coherent blocks of debris after the catastrophic 
collisional disruption of a parent body~\citep{mich13}.  Repeated fission and impact of the components by any of these
scenarios causes a so-called contact-binary cycle~\citep{sche07}.

The effect of a nonspherical component can shift the locations of the zero-energy and stability limits for specific binary
systems by up to a few tens of percent in scaled separation compared to the two-sphere case.  Despite this, the 
zero-energy limit crosses the contact limit near $v=0.15-0.2$ and $v=0.80-0.85$, similar to the ranges found 
by~\citet{sche09}, indicating that, upon fissioning from the parent body, small secondaries are at risk of becoming 
gravitationally unbound from the system.  The resulting asteroid pair could then have a mass ratio of less than one to four, 
compared to one to five in the two-sphere case, which is consistent with observation~\citep{prav10}.  \citet{sche09} 
considers extending to higher-order expansions of the gravitational potential in the future.  We find that the approximate 
zero-energy and stability limits presented by~\citet{sche09} in the oblate case are accurate to better than 1\% for 
$\alpha>0.5$ compared to our solution that uses the exact gravitational potential.  In the prolate case, the differences 
can grow to $1-2\%$ for $\beta=0.75$ and $5-10\%$ for $\beta=0.5$, suggesting that the~\citet{sche09} approximation 
is reliable for ellipsoids except in the most extreme cases of nonspherical shapes.


\section*{Acknowledgments}
\label{sec:ack}

\noindent
This material is based upon work supported by the National Aeronautics and Space Administration under Grant No. 
NNX12AF24G issued through the Near-Earth Object Observations Program.


\clearpage

\bibliographystyle{icarus}
\bibliography{TaylorMargot-Nonspherical}

\begin{thebibliography}{}

\bibitem[{Bellerose} and {Scheeres}(2008){Bellerose} and {Scheeres}]{bell08}
{Bellerose}, J., {Scheeres}, D.~J., 2008.
\newblock {Energy and stability in the full two-body problem}.
\newblock Cel. Mech. Dyn. Astron.~100, 63--91.

\bibitem[{Descamps} and {Marchis}(2008){Descamps} and {Marchis}]{desc08}
{Descamps}, P., {Marchis}, F., 2008.
\newblock {Angular momentum of binary asteroids: Implications for their
  possible origin}.
\newblock Icarus~193, 74--84.

\bibitem[{Jacobson} and {Scheeres}(2011){Jacobson} and {Scheeres}]{jaco11}
{Jacobson}, S.~A., {Scheeres}, D.~J., 2011.
\newblock {Dynamics of rotationally fissioned asteroids: Source of observed
  small asteroid systems}.
\newblock Icarus~214, 161--178.

\bibitem[{Magri} et~al.(2011){Magri}, {Howell}, {Nolan}, {Taylor},
  {Fern{\'a}ndez}, {Mueller}, {Vervack}, {Benner}, {Giorgini}, {Ostro},
  {Scheeres}, {Hicks}, {Rhoades}, {Somers}, {Gaftonyuk}, {Kouprianov},
  {Krugly}, {Molotov}, {Busch}, {Margot}, {Benishek}, {Protitch-Benishek},
  {Gal{\'a}d}, {Higgins}, {Ku{\v s}nir{\'a}k}, and {Pray}]{magr11}
{Magri}, C., and 25 colleagues, 2011.
\newblock {Radar and photometric observations and shape modeling of contact
  binary near-Earth asteroid (8567) 1996 HW1}.
\newblock Icarus~214, 210--227.

\bibitem[{Margot} et~al.(2006){Margot}, {Pravec}, {Nolan}, {Howell}, {Benner},
  {Giorgini}, F., {Ostro}, {Slade}, {Magri}, {Taylor}, {Nicholson}, and
  {Campbell}]{marg06iau}
{Margot}, J.~L., and 12 colleagues, 2006.
\newblock Hermes as an exceptional case among binary {near-Earth} asteroids.
\newblock IAU Gen. Assem.~236, (Abstract \#S236--35).

\bibitem[{Michel} and {Richardson}(2013){Michel} and {Richardson}]{mich13}
{Michel}, P., {Richardson}, D.~C., 2013.
\newblock {Collision and gravitational reaccumulation: Possible formation
  mechanism of the asteroid Itokawa}.
\newblock Astron. \& Astrophys.~554, L1, 1--4.

\bibitem[{Ostro} et~al.(2006){Ostro}, {Margot}, {Benner}, {Giorgini},
  {Scheeres}, {Fahnestock}, {Broschart}, {Bellerose}, {Nolan}, {Magri},
  {Pravec}, {Scheirich}, {Rose}, {Jurgens}, {De Jong}, and {Suzuki}]{ostr06}
{Ostro}, S.~J., and 15 colleagues, 2006.
\newblock {Radar imaging of binary near-Earth asteroid (66391) 1999 KW4}.
\newblock Science~314, 1276--1280.

\bibitem[{Pravec} et~al.(2010){Pravec}, {Vokrouhlick{\'y}}, {Polishook},
  {Scheeres}, {Harris}, {Gal{\'a}d}, {Vaduvescu}, {Pozo}, {Barr}, {Longa},
  {Vachier}, {Colas}, {Pray}, {Pollock}, {Reichart}, {Ivarsen}, {Haislip},
  {Lacluyze}, {Ku{\v s}nir{\'a}k}, {Henych}, {Marchis}, {Macomber}, {Jacobson},
  {Krugly}, {Sergeev}, and {Leroy}]{prav10}
{Pravec}, P., and 25 colleagues, 2010.
\newblock {Formation of asteroid pairs by rotational fission}.
\newblock Nature~466, 1085--1088.

\bibitem[{Scheeres}(1994){Scheeres}]{sche94}
{Scheeres}, D.~J., 1994.
\newblock {Dynamics about uniformly rotating triaxial ellipsoids: Applications
  to asteroids}.
\newblock Icarus~110, 225--238.

\bibitem[{Scheeres}(2006){Scheeres}]{sche06}
{Scheeres}, D.~J., 2006.
\newblock {Relative equilibria for general gravity fields in the
  sphere-restricted full 2-body problem}.
\newblock Celest. Mech. Dyn. Astron.~94, 317--349.

\bibitem[{Scheeres}(2007){Scheeres}]{sche07}
{Scheeres}, D.~J., 2007.
\newblock {Rotational fission of contact binary asteroids}.
\newblock Icarus~189, 370--385.

\bibitem[{Scheeres}(2009){Scheeres}]{sche09}
{Scheeres}, D.~J., 2009.
\newblock {Stability of the planar full 2-body problem}.
\newblock Cel. Mech. Dyn. Astron.~104, 103--128.

\bibitem[{Taylor} and {Margot}(2011){Taylor} and {Margot}]{tayl11material}
{Taylor}, P.~A., {Margot}, J.~L., 2011.
\newblock {Binary asteroid systems: Tidal end states and estimates of material
  properties}.
\newblock Icarus~212, 661--676.

\end{thebibliography}

\end{document}